\documentclass[twocolumn,showpacs,preprintnumbers,amsmath,amssymb]{revtex4}

\usepackage{graphicx}
\usepackage{dcolumn}
\usepackage{bm}

\begin{document}

\title{Depinning and creep motion in glass states of flux lines}

\author{Meng-Bo Luo\(^{\dagger,\ddagger}\) and Xiao Hu\(^{\dagger}\)}

\affiliation{ \(^{\dagger}\)Computational Materials Science Center,
National Institute for Materials Science, Tsukuba 305-0047, Japan
\\ \(^{\ddagger}\)Department of Physics, Zhejiang University, Hangzhou 310027,
China}

\date{\today}

\begin{abstract}
Using dynamical computer simulation we have investigated 
vortex matters in glass states. A genuine continuous 
depinning transition is observed at zero temperature, which 
also governs the low-temperature creep motion. With the notion
of scaling, we evaluate in high accuracy critical exponents and
scaling functions; we observe a non-Arrhenius creep motion 
for weak collective pinning where Bragg glass (BrG) is stabilized at equilibrium, 
while for strong pinning the well-known Arrhenius law is recovered. In both cases, 
a sharp crossover takes place between depinning and creep at low temperatures. 
The possible relation between the present results
and a recent experimental observation of 
a second-order like phase boundary inside the BrG phase is discussed.

\end{abstract}


\maketitle

\noindent {\it Introduction --} 
Depending on the strength of the random pinning force, the equilibrium state of the
flux lines at low enough temperature can be either Bragg glass (BrG) 
\cite{BG_Nattermann,BG_GL}, where quasi long-range lattice order still survives,
or vortex glass (VG)  \cite{VG} as random as the liquid phase.
The competition between the vortex repulsion and the random 
pin potential builds up a highly nontrivial energy landscape, which 
manifests itself drastically in dynamics, e.g. vortex motion under current driving. 
Since the proposal of collective pinning theory of vortex lines by Larkin 
and Ovchinnikov \cite{LO},  theoretical understanding for the nonlinear 
dynamics response has been advanced  
\cite{Ioffe1987,Nattermann1987,Feigel'man1989,BG_Nattermann,Blatter_ROM}.
The functional renormalization group (FRG) has also been formulated
\cite{FRG}. For recent review articles see 
\cite{Review_Giamarchi,Review_Nattermann2000,Review_Nattermann2004}.

The current-driven flux lines compose a unique system of \(D=3\) dimensions 
in the internal space, equal to the dimension \(d=3\) of the space where the
system is embedded, and \(N=2\) components of displacement vector.
A full FRG treatment is still not available for 
\(N>1\). We try to address the issue by computer simulations, with the hope that 
useful insights complimentary to the previous works can be provided. It is recalled that
computer simulations for a domain wall in a plane with
\(N=1\), \(D=1\) and \(d=2\) were reported recently \cite{Kolton}. 

On the other hand, a new experimental finding of a second-order like phase boundary
in \(H-T\) phase diagram was reported very recently, which intersects the first-order phase
boundary associated with the melting transition of the BrG \cite{Beidenkopf}.
It certainly renews interests on corresponding variations in dynamical properties
inside the BrG phase; the new phase transition was discussed in terms
of replica-symmetry breaking \cite{Rosenstein_Li},  which is perhaps best captured by 
dynamical responses. 

The main results of the present work are summarized as follows:
Based on the simulation results, we have derived a scaling relation among
the velocity, force and temperature with two universal exponents for
vortex motions around the zero-temperature depinning force. From the 
exponents and the scaling curve, the Arrhenius law for the creep motion
with a linearly suppressed energy barrier appears for strong pinning strength. A 
non-Arrhenius type creep motion is observed at weak collective pinning for which 
the equilibrium state is a BrG. A sharp crossover between the depinning 
and creep motion is also observed.

\vspace{3mm}

\noindent {\it Model and simulation details --} We consider a superconductor
of layered structure with magnetic field perpendicular to the
layers. The model system is a stack of
superconducting planes of thickness $d$ with interlayer space $s$.
Each plane contains $N_v$ pancake vortices (PV) 
and $N_p$ quenched pins. The overdamped equation of motion of the
\(i\)th pancake at position ${\bf r}_i$ is 
\cite{Ryu,Reichhardt,Otterlo00,Olive,Chen_Hu}

\begin{equation}
   \eta \dot{{\bf r}}_i = -\sum\limits_{j\neq i} \nabla_i
   U^{VV}({\bf r}_{ij})- \sum\limits_{p} \nabla_i
   U^{VP}({\bf r}_{ip}) + {\bf F}_L + {\bf F}_{th} .
\end{equation}

\noindent Here $\eta$ is the viscosity coefficient. The
vortex-vortex interaction contains two parts: an
intraplane PV-PV pairwise repulsion given by the modified Bessel
function $U^{VV}(\rho_{ij},z_{ij}=0) = d\epsilon_0
K_0(\rho_{ij}/\lambda_{ab})$, and an interplane
attraction between PVs in adjacent layers given by
$U^{VV}(\rho_{ij},z_{ij}=s)=(s\epsilon_0 /\pi) [1+\ln
(\lambda_{ab}/s)][(\rho_{ij}/2r_g)^2-1]$ for $\rho_{ij} \leq 2r_g$
and $U^{VV}(\rho_{ij},z_{ij}=s)=(s\epsilon_0 /\pi) [1+\ln
(\lambda_{ab}/s)][\rho_{ij}/r_g-2]$ otherwise, where
$\epsilon_0=\phi_0^2/2\pi\mu_0\lambda_{ab}^2$ with $\lambda_{ab}$
the in-plane magnetic penetration depth, $r_g = \gamma \xi_{ab}$
with $\xi_{ab}$ the in-plane coherence length and $\gamma$ the
anisotropy \cite{Olive,Ryu}. Here, $\rho_{ij}$ is the in-plane
component of position vector ${\bf r}_{ij}$ between \(i\)th and
\(j\)th PVs and $z_{ij}$ is
that along layer normal. The pinning potential is given by
$U^{VP}(\rho_{ip}) = -\alpha A_p\exp[-(\rho_{ip}/R_p)^2]$, where
$A_p = (\epsilon_0 d/4)\ln [1 + (R_p^2/2\xi_{ab}^2)]$ and $\alpha$
is the dimensionless pinning strength. Finally, ${\bf
F}_L$ is the uniform Lorentz force, and ${\bf F}_{th}$ is the
thermal noise force with zero mean and a correlator $\langle
F^\alpha_{th}(z,t) F^\beta_{th}(z^\prime,t^\prime)\rangle =2\eta
T\delta^{\alpha\beta}\delta (z-z^{\prime })\delta (t-t^{\prime })$
with $\alpha, \beta = x, y$.  We consider a material
similar to Bi\(_2\)Sr\(_2\)CaCu\(_2\)O\(_8\) with
$\kappa=\lambda_{ab}/\xi_{ab}=90$, $\gamma = 100$,
and $d=2.83\times 10^{-3}\lambda_{ab}$, $s=8.33\times
10^{-3}\lambda_{ab}$, with $R_p = 0.22\lambda_{ab}$ \cite{Olive}. 
Below the units for length, energy, termperature, force and time
are taken as $\lambda_{ab}$,  $d\epsilon_0$, $d\epsilon_0/k_B$,    
 $d\epsilon_0/\lambda_{ab}$,  and 
$\eta \lambda^2_{ab}/d\epsilon_0$.
Pinning strength is supposed to be uniform. 
Periodic boundary conditions are put in all the three
directions. The results shown below are for \(N_v=180\),
\(N_p=900\), \(N_z=20\), \(L_{xy}=30\).
The magnetic field is roughly \(B\simeq 100 G\) if we take 
$\lambda_{ab} = 2000\AA$ \cite{Zeldov}. 
The equation is integrated by the 2nd order Rugen-Kutter algorithm with
\(\Delta t=0.002\sim 0.01\). All data presented below are
the average over 10 samples with different randomly distributed pins.
Fixing the magnetic field,  we have performed simulations for  
\(L_{xy}=20\) and \(40\), and make sure 
that finite-size effects are negligible to the main, universal results.
For pinning strength, we choose \(\alpha=0.2\) and \(\alpha=0.05\),
for which the maximal curvatures of the pinning potential 
are 4.87 and 1.22, above and below the tilt modulus of the 
flux lines \(\bar C\simeq 1.47\); they are therefore expected 
to fall into the strong and weak collective
pinning regimes respectively \cite{Labusch,Blatter2004}.

\vspace{3mm}

\begin{figure}[t]
\vspace{6cm}
\includegraphics{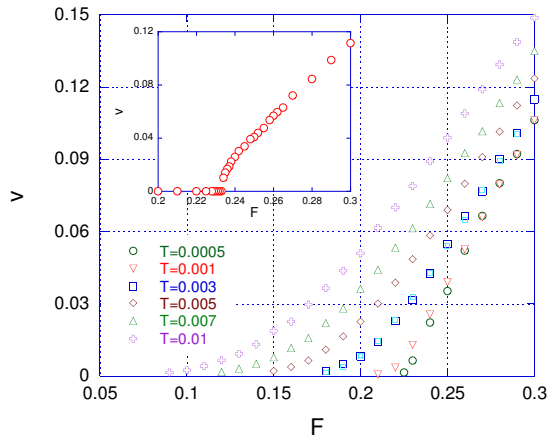}
\caption{\(v-F\) characteristics for \(\alpha=0.2\). Data for \(T=0.003\)
obtained for \(L_{xy}=40\lambda_{ab}\) are also included (light-blue squares). 
Inset: data for zero temperature. Error bars are smaller than the size of symbols.}
\label{fig:vfa02}
\end{figure}

\noindent {\it Strong pinning --}
Let us begin with the strong pinning case of \(\alpha=0.2\). The
equilibrium state is a pinned solid (VG) as random as liquid 
seen from the structure factor.
The \(v-F\) characteristics at low temperatures are depicted in 
Fig. \ref{fig:vfa02} (\(T=0\) in inset). A continuous depinning transition 
is observed at \(T=0\) with a unique depinning force \cite{Middleton}, 
which can be described by
\(v\simeq A(F/F_{c0}-1)^{\beta}\) with \(F_{c0}\simeq 0.231\pm 0.002\) and
\(\beta \simeq 0.74\pm 0.02\).  For \(0<T<0.0005\),
upward-convex \(v-F\) characteristics 
are observed down to \(F_{c0}\); below \(F_{c0}\) there
is an extremely small tail which is hard to see in the present scale.

The sharp depinning transition is clearly rounded by finite temperatures.
In order to explore the critical properties of the \(v-F\) characteristics 
at finite temperatures, we postulate the following
scaling ansatz \cite{Fisher,Middleton2,FMdomain}

\begin{equation}
v(T,F)=T^{1/\delta}S(T^{-1/\beta\delta}f)
\label{eqn:sceqaution}
\end{equation}

\noindent with \(f=1-F_{c0}/F\) \cite{note} and the scaling function 
\(S(x)\) regular at \(x=0\) and \(S(x)\rightarrow x^\beta\) as 
\(x\rightarrow +\infty\).

\begin{figure}[t]
\vspace{6cm} 
\includegraphics{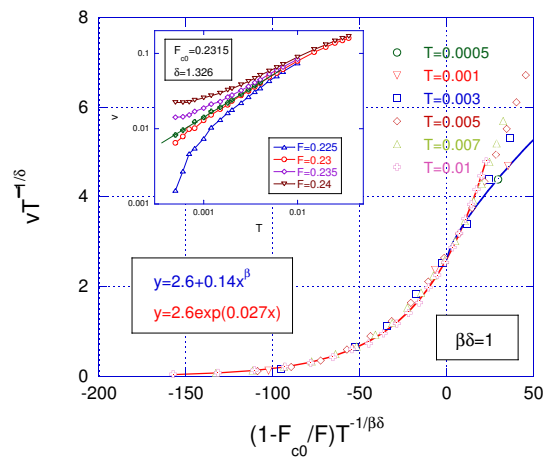}
 \caption{Scaling plot for the data in Fig.\ref{fig:vfa02}. Inset: Temperature
dependence of velocity for several forces around \(F_{c0}\).}
\label{fig:scplota02}
\end{figure}

The critical force can be determined using the property
\(v(T,F=F_{c0})= S(0) T^{1/\delta}\) implied in Eq. (\ref{eqn:sceqaution})
(see also \cite{NPV2001}).
As shown in the inset of Fig. 2, we evaluate \(F_{c0}=0.2315\pm 0.0013\),
and meanwhile from the slope \(1/\delta=0.754\pm 0.010\). 
We then perform the scaling plot using the
scaled variables \(vT^{-1/\delta}\) and
\((1-F_{c0}/F)T^{-1/\beta\delta}\); the best collapsing of data to a
single scaling curve is achieved when \(\beta\simeq
1/\delta=0.754\), thus determining the exponent \(\beta\). The 
values of \(F_{c0}\) and \(\beta\) estimated from data at finite
temperatures via the scaling analysis are consistent with those
derived from \(T=0\), which can be taken as an evidence for
the existence of scaling.  Fitting the scaling curve, we obtain
\(S(x)=0.14x^\beta+2.6\) for \(x>0\) and \(S(x)=2.6\exp(0.027x)\)
for \(x<0\); the former covers the continuous
depinning transition at \(T=0\); the latter, combined with the
relation \(\beta\delta=1\), indicates that the motion at
low temperatures and forces below \(F_{c0}\) is
well described by the Arrhenius law, and that the energy barrier
disappears linearly when the force is ramped up to \(F_{c0}\); the
bare energy barrier is \(U_c=0.027\).

The same exponents and similar scaling behaviors are available for
\(\alpha=0.4\); therefore, the above properties are 
universal for strong pinning case. The creep law derived above 
confirms the {\it a priori} assumption in the Anderson-Kim
theory \cite{Anderson_Kim}, and coincides with the FRG results in Ref. \cite{Muller_FRG} 
for a domain wall (\(N=1\)). Our results are also consistent 
with Ref. \cite{NPV2001} provided \(\theta=1\). 

Deviations from the scaling curve are observed for
(a) \(F_{c0}/2<F<F_{c0}\) at \(T\ge 0.015\simeq U_c/2\), 
due to extra thermal deformation of flux-line lattice; (b)
\(F<F_{c0}/2\) which may be governed by the zero-force limit;
and (c) \(F\gg F_{c0}\) the flux-flow regime.
 
\vspace{3mm}

\begin{figure}[t]
\vspace{6cm}
\includegraphics{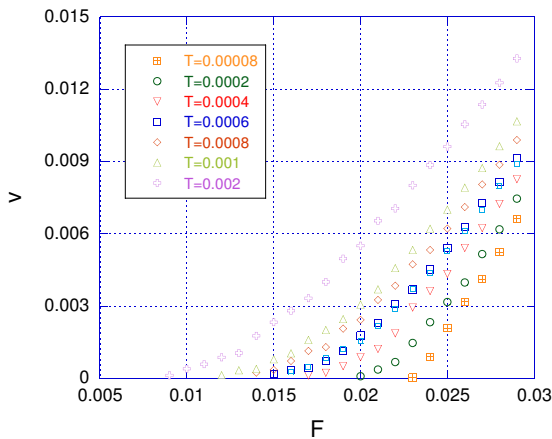}
\caption{\(v-F\) characteristics for \(\alpha=0.05\). Data for \(T=0.0006\)
obtained for \(L_{xy}=40\lambda_{ab}\) are also included (light-blue squares).}
\label{fig:vfa005}
\end{figure}

\begin{figure}[t]
\vspace{6.5cm}
\includegraphics{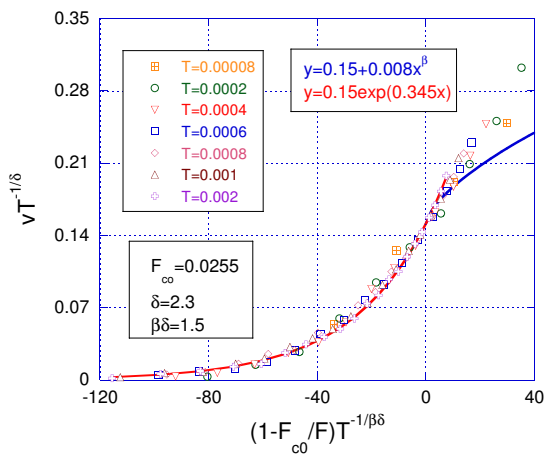}
\caption{Scaling plot for data in Fig. \ref{fig:vfa005}.}
\label{fig:scplota005}
\end{figure}

\noindent {\it Weak collective pinning --}
We have performed the same simulations for \(\alpha=0.05\), which
falls into the weak collective pinning regime. The equilibrium state is a BrG, with
the melting temperature \(T_m\simeq 0.077\) above which the quasi long-range
order is suppressed by thermal fluctuations.  Looking at the velocity-velocity
correlation function, we find in this case that the moving system shows 
good temporal and spatial orders characterized by Bragg peaks \cite{Balents,MBG} 
(No moving smectic \cite{MSmectic} was observed for the present parameters.)
The \(v-F\) characteristics are shown in Fig. \ref{fig:vfa005},
where similar to the strong pinning case, a sharp crossover between the depinning 
and creep motion takes place at \(F_{c0}\) for \(0<T\le 0.00008\). The scaling plot is depicted 
in Fig. \ref{fig:scplota005}. The exponents are estimated as \(\delta=2.3\pm 0.1\)
and \(\beta=0.65\pm 0.01\), which are different from those for strong
pinning case. The product of the two exponents
\(\beta\delta=3/2\) deviates from unity, indicating a nonlinear scaling
relation between temperature and force deviation from \(F_{c0}\).

Nonlinear scaling relations have been found in CDW systems \cite{Middleton2}. 
This behavior can be captured by an effective 
potential of linear and cubic terms of displacement, with a small energy 
barrier \cite{Middleton2,Review_Nattermann2004}.  The linear scaling relation
observed for strong pinning force corresponds to an effective potential
of linear, quadratic and quartic terms of displacement, with a large energy
barrier \cite{STM}.  It is observed in our simulations that the system with 
weak bare pinning strength experiences smaller energy barriers compared with 
that of strong bare pinning strength, even at the \textit{same} relative force-deviation
from the critical values;  an analytic derivation of the
effective energy barrier, however, is not an easy task \cite{Review_Nattermann2004}.
While a full picture remains to be developed, we notice that, first, vortex motions
mimic nucleation processes in first-order phase transitions: 
the weak/strong pinning case corresponds to a system locates at the 
spinodal/coexistence curve \cite{STM}; secondly, under weak and strong pinning vortices
behave similarly to CDW \cite{Middleton2} and domain wall \cite{Muller_FRG} respectively,
due to the different ranges of correlation.

The scaling curve in Fig. \ref{fig:scplota005} is fitted well by  
\(S(x)=0.15\exp(0.0345x)\), which, with the nonlinear scaling variable,
indicates a non-Arrhenius creep motion for the weak pinning case. To the 
best of our knowledge, this behavior has not been reported so far.  On a 
phenomenological level, it can be captured with an appropriate
correlator of energy barriers \cite{LeDoussal_Vinokur}. 

\vspace{-5mm}

\begin{figure}[t]
\begin{center}
\vspace{-0.5cm} 
\includegraphics[width=6.3cm]{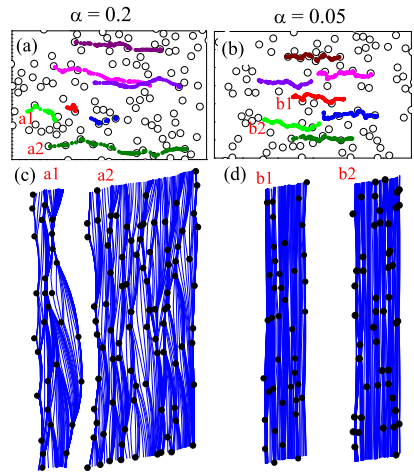}
\vspace{-0.8cm}
\caption{Vortex motions at \(T=0\). (a) and (b):
Trajectories of 7 nearest-neighboring vortices on one layer for $\alpha =0.2$
at $F=0.24$ and $\alpha=0.05$ at $F=0.027$ respectively. 
(c) and (d): Trajectories of the two flux lines indicated in (a) and (b). 
The total times are  330 for $\alpha =0.2$, and 1200 for $\alpha =0.05$,
such that flux lines in both systems move roughly the same
distance \(4\lambda\sim 1.7a_0\) in average. Open circles in (a) and (b) 
represent the positions of pins, while the filled circles in (c) and (d) 
represent pins that pin the flux lines. }
\label{fig:motion}
\end{center}
\end{figure}

\vspace{6mm}

\noindent {\it "Microscopic" vortex motion--} 
We have examined the {\it microscopic} motions of flux lines.
As detailed in Fig. \ref{fig:motion}  for forces slightly above \(F_{c0}\)
and at zero temperature (similar results have been obtained 
for \(F<F_{c0}\) at finite temperatures), flux lines in the weak pinning 
case move homogeneously in an intermediate time scale, which guarantees 
the moving BrG (mBrG) order \cite{MBG}; 
flux lines under strong pinning move in an inhomogenous 
way, and thus additional dislocations are induced during the motion. 
Even for the case of mBrG, velocity fluctuates for time being and between 
different vortices, as can be seen in Fig. \ref{fig:motion}d.
Namely in a shorter time scale, the motion of vortices in weak
pinning case is also random: 
Flux lines exhibit an intermittent pattern of motion, {\it i.e.} a portion of
flux lines move while others are almost motionless at a given time window,
and in the next time window, similar situation occurs with the moving
vortices found in different regions. The motion of the total system
is thus an accumulation of individual, random movements, in which
creep events dominate and the random pinning plays an essential role.
As shown in Figs. \ref{fig:motion}c and d the shape of vortex lines 
upon depinning depends on pin strength \cite{Blatter2004}.

\vspace{3mm}

\noindent {\it Disscussions--}
The strong pinning \(\alpha=0.2\) and weak collective
pinning \(\alpha=0.05\) according to the Labusch criterion \cite{Labusch,Blatter2004}
result in vortex glass and Bragg glass respectively. But generally the Labusch criterion 
does not coincide with the phase boundary of Bragg glass. Further study is expected 
to resolve this point in detail.

As can been seen in Figs. \ref{fig:motion}a and b, during the creep motion vortices 
displace frequently into the direction transverse to the force and the 
averaged velocity. In this way a bypath of lower energy can be found 
which assists vortices to avoid otherwise high energy barriers. This may
affect the averaged velocity significantly since the time spent
in large energy barriers is huge, and is to be taken into 
account carefully in theoretical treatments.

Beside the genuine, zero-temperature depinning transition, we observe  
a quite steep crossover between the depinning and creep motions
at low temperatures (\(T\le 0.0005\) for \(\alpha=0.2\) and \(T\le 0.00008\) 
for \(\alpha=0.05\)). The
velocity tail below the crossover is very small as such it
looked like a phase transition in simulations (and perhaps also in experiments). 
The crossover, however, loses its sharpness when
temperature increases but still far below the melting temperature, which
might be not quite consistent with experimental phase boundary 
\(H_g(T)\) \cite{Beidenkopf}.  The possibility of a transition (or sharp crossover) between
the depining and flux flow regimes remains to be investigated within the
present formalism, taking into account the shaking magnetic field.


\noindent {\it Acknowledgements --}
We are pleased to acknowledge T. Nattermann, D. Genshkenbein, G. Blatter,
B. Rosenstein, V. Vinokour and D. P. Li for discussions. 
Calculations were performed on SR11000 (HITACHI) in NIMS. 
X. H. is supported by Grant-in-Aid for Scientific Research (C) No. 18540360
of JSPS, and project ITSNEM of China Academy of Science.


\begin{thebibliography}{99}
\bibitem{note} The usually adopted quantity \(f'=1-F/F_{c0}\) gives the same
exponents, but the scaling regime is much narrower.
\end{thebibliography}
\end{document}